\documentstyle[bo99,epsfig]{article}

\newcommand{\degree} {\mbox{$^\circ$}}
\newcommand{\etal} {{\em et al.}}

\def\arcsec{\hbox{$^{\prime\prime}$}}

\title{The USA X-ray Timing Experiment}

\author{P. S. Ray$^1$, K.S. Wood$^1$, G. Fritz$^1$, P. Hertz$^1$,
M. Kowalski$^1$, W.N. Johnson$^1$, M.N. Lovellette$^1$,
M.T. Wolff$^1$, D. Yentis$^1$, R. M. Bandyopadhyay$^2$,
E.D. Bloom$^3$, B. Giebels$^3$, G.Godfrey$^3$, K. Reilly$^{3,4}$, P. Saz
Parkinson$^{3,4}$, G. Shabad$^{3,4}$, P. Michelson$^4$, M. Roberts$^4$,
D.A. Leahy$^5$, L. Cominsky$^6$, J. Scargle$^7$, J. Beall$^8$,
D. Chakrabarty$^9$, Y. Kim$^{10}$}

\affil{1) E. O. Hulburt Center for Space Research, Naval Research Laboratory, 
2) NRC Research Associate, 
3) Stanford Linear Accelerator Center,
4) Stanford University,
5) University of Calgary,
6) Sonoma State University,
7) NASA Ames,
8) Saint John's College,
9) MIT,
10) Saddleback College}

\begin{document}

\maketitle

\begin{abstract}
The USA Experiment is a new X-ray timing experiment with large
collecting area and microsecond time resolution capable of conducting
a broad program of studies of galactic X-ray binaries.  USA is one of
nine experiments aboard the Advanced Research and Global Observation
Satellite which was launched February 23, 1999.  USA is a collimated
proportional counter X-ray telescope with about 1000 cm$^2$ of
effective area per detector with two detectors sensitive to photons
in the energy range 1--15 keV.  A unique feature of USA is that
photon events are time tagged by reference to an onboard GPS receiver
allowing precise absolute time and location determination.  
We will present an overview of the USA instrument, capabilities, and
scientific observing plan as well as the current status of the
instrument.
\end{abstract}

\section{Introduction}

The Unconventional Stellar Aspect (USA) Experiment is a low-cost X-ray
timing experiment with the dual purpose of timing X-ray binary systems
and exploration of applications of X-ray sensor technology.  USA was
launched on February 23, 1999 on the Advanced Research and Global
Observation Satellite (ARGOS).  It is a reflight of two proportional
counter X-ray detectors that performed excellently on the NASA
Spartan-1 mission (Kowalski \etal\ 1993).  The primary targets are
bright Galactic X-ray binaries that are used simultaneously for both
scientific and applied objectives. X-ray photon event times are
measured to high precision using the GPS receiver on ARGOS.  USA has
the effective area, precise timing ability, and data throughput
capability to probe these sources at the timescales of processes near
neutron star surfaces or the innermost stable orbits around black
holes.  A second objective of the experiment is to conduct experiments
involving applied uses of X-ray detectors in space and with
reliable computing in space.  These will not be discussed here but
descriptions are available elsewhere (Wood 1993). 

Key characteristics of the experiment and mission that facilitate this
overall program include (i) a mission concept that allows long
observing times on bright X-ray objects, (ii) large-area detectors
with high time resolution capability (effective area: 2000 cm$^2$;
telemetry: 40 kbps, with 128 kbps available for short periods; 2
$\mu$s time resolution), (iii) good low energy response (down to 1
keV), and (iv) a high flexibility in data handling.  Other special
features include absolute time-tagging (to 2 $\mu$s) using a GPS
receiver.

\section{Scientific Program}

The principal targets for USA are X-ray binaries whose X-ray emitting
members are neutron stars, black holes, or white dwarfs.  Study of
physical processes in these systems have been among the main thrusts
of X-ray astronomy since the founding of the field.  Today it remains
true that many of the most important results on these systems are
found by studying their X-ray variability, and the push to shorter
(millisecond) timescales is proving highly fruitful.  If the source is
bright ($>$ milliCrabs) such short timescales are more readily reached
with non-imaging instruments having large collecting apertures than
with imaging instruments.  Physics issues studied in these sources are
generally related to the fact that parameters such as magnetic field
strength, mass and energy densities, and gravitational fields reach
extreme values, hence providing the preferred testing grounds for
physical theories.  X-ray timing is a cornerstone of relativistic
astrophysics.

USA, in turn, is one of the two main resources at the present epoch
for X-ray timing experiments, the other being the Proportional Counter
Array (PCA) on RXTE.  USA has its own special areas of emphasis, one
of which is its observing plan. Present plans call for the observation
of about 30 primary targets, with each being observed for about 1
month over a nominal mission life of 3 years; selected targets will be
observed for shorter periods of time.  Sources observed to date
(through 31 August 1999) include Cyg~X-1 (700~ks on target), Aql~X-1
(100~ks), Cen~X-3 (65~ks), X1630-472 (60~ks), Cyg~X-2 (50~ks),
X1636-536 (45~ks), GX~1+4 (40~ks), 1E2259+586 (40~ks), X1820-30
(40~ks), X1630-472 (35~ks), 1E1048.1-5937 (30~ks), and GRS~1915+105
(25~ks).  The total time on each source is typically scheduled as a
number of $\sim$1~ks observations distributed over weeks or
months. Simultaneous observation with other observatories, such as the
Compton Gamma Ray Observatory and Rossi X-ray Timing Explorer, and
with ground based telescopes are also being undertaken.

Figure~1 shows two sample light curves taken with USA.  The first is
an X-ray burst from the burster X1735-444, and the second is an
observation of a flaring state of the Galactic microquasar
GRS1915+105.  In 1735-44 the instrument is on the source throughout
the interval displayed while in GRS 1915+105 the steep rise at the
beginning of the plot is the instrument slewing onto the source; 
the earliest seconds represent the background for this observation.

\begin{figure}[t]
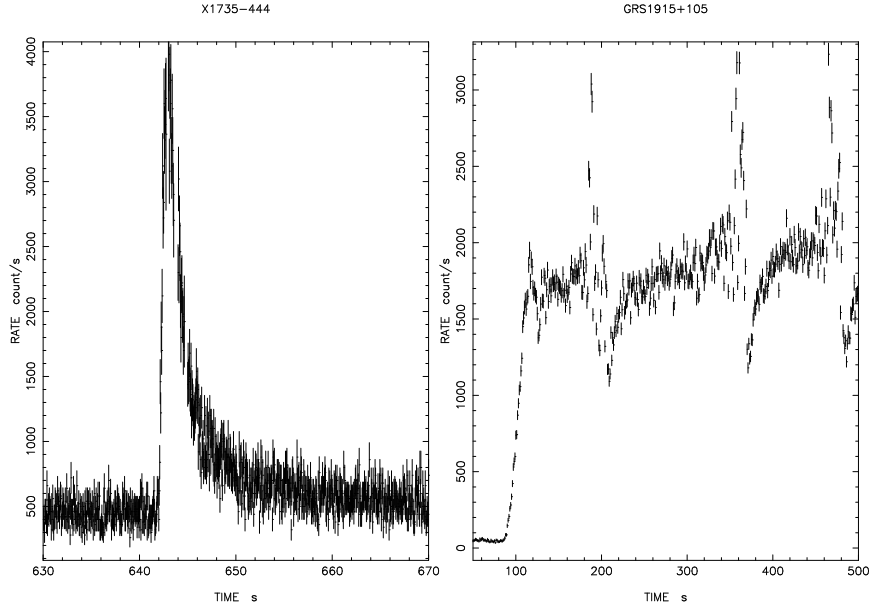

\begin{center}
\mbox{\psfig{file=X1735-444b.ps, width=2.25in}\psfig{file=GRS1915b.ps,
width=2.25in}}
\caption{Left: A burst from X1735-444.  Right: An observation of
GRS1915+105}
\end{center}
\end{figure}

\subsection{Low-Mass X-ray Binaries}

The special importance of the low mass X-ray binaries (LMXBs) arises
from their comparatively weak magnetic fields which allows the disk to
penetrate very close to the star.  This gives rise to fast timing effects that
can be used to probe the extreme conditions in the neutron star
vicinity. Major gains in the understanding of these phenomena have
been made since the launch of the Rossi X-ray Timing Explorer (RXTE)
in late 1995.  High frequency quasiperiodic oscillations (QPOs) and
short strings of coherent pulsations during bursts have been used to
argue convincingly that effects associated with inner disk edges and
the innermost stable orbits predicted by General Relativity are being
seen (van der Klis 1998). Another milestone is the establishment of
the spinup evolution of neutron stars through the discovery of the
first accretion powered millisecond pulsar (SAX J1808.4-3658).

USA will make further contributions to the study of LMXBs with the
application of its unique strengths.  In some cases this will mean
exploiting the ability to dedicate large blocks of time to a key
source, e.g., to refine understanding of SAX J1808.4-3658 or to
observe transitions between modes or states.  Significant time is
being devoted to searches for coherent periods, both on and off
bursts.  Off burst work is carried out using coherence recovery
searches for periods.  Observations can also be carried out in various
ways to detect or refine orbital periods in LMXBs.  Overall, LMXBs are
sources that stand to bring major rewards including advances in
understanding the role of General Relativity in the dynamics of inner
disk regions, but past experience has also shown that these rewards
are achieved only through major investments of observing time and
analysis, chiefly because of the elusiveness and short timescale of
the spin periods.

\subsection{High-Mass X-ray Binaries}

USA will also accumulate significant time on a number of high-mass
X-ray binaries.  Many of these systems have accretion rates that far
exceed the Eddington limit locally in the accretion column.  This
means that radiation pressure has a significant influence on the flow.
Recently Jernigan \etal\ (1999) reported the discovery of Photon
Bubble Oscillations (PBOs) in Cen X-3.  USA will observe this source
and other bright HMXBs to help characterize their high frequency
power spectra independently from RXTE.  Outstanding puzzles in these
systems are the details of angular momentum transfer from the disk to
the star (including possible reversals of the sense of disk rotation)
and understanding in detail the photohydrodymics of the accretion
column in which the super Eddington accretion funneled flow is
converted to the observed X-ray emission.  Bright binary pulsars such
as Her X-1, Cen X-3, and Vel X-1 will be observed to gain insight into
these issues. Monitoring over both short and long time periods allows
the correlation between period, period derivative, and luminosity to
be probed which addresses the angular momentum transfer issue.  USA
observes at significantly lower energies than the BATSE instrument on
CGRO, which has gathered much of the data on this topic in recent
years.  

\subsection{Black Hole Candidates}

USA will pursue several investigations into the nature of black holes.
Chief among these are (1) The characterization of high frequency ($\nu
> 1$~kHz) variability in Cyg X-1.  USA has begun a long term study of
Cyg X-1 in which more than 1 Ms of exposure will be accumulated on Cyg
X-1 in each of the spectral states it exibits during the USA three
year mission.  Using techniques developed for calibrating high
frequency systematic effects in HEAO A-1 and RXTE (Chaput \etal\
1999), we will constrain the sub-millisecond variability of Cyg X-1.
(2) Simultaneous X-ray and infrared observations of the galactic
microquasar GRS~1915+105.  USA will, for the first time, determine the
1--3~keV behavior of this interesting source.  Figure 1 does not begin
to convey the range of modulation patterns seen in this object.  A new
window at lower energies from USA, gathered in simultaneity with other
space-based and ground facilities, may contribute to modeling the
fluid dynamical processes near the black hole.  The model that can
account for the wealth of variability effects seen in GRS 1915+105 may
bring us closer to understanding how plasma behaves near black holes,
including the relativistic effects on orbits.

\subsection{Other Sources}

The Anomalous X-ray Pulsars (AXPs) appear to be a distinct subpopulation
whose observed emission is powered by spin-down.  Timing these pulsars
over a period of years will place strong constraints on their nature
(whether they are magnetars or accreting neutron stars) and emission
mechanism.  Subtle effects of (currently unsuspected) binary
companions may show up or else derivatives that provide insight into
source dynamics may be measured.  There are several good candidates
and the result will not necessarily be the same in all instances.
Long-term monitoring of AXPs is made feasible by the soft response of
USA and the good absolute timing.  Rotation-powered (radio) pulsars
are also observed by USA to validate the time transfer between USA and
RXTE, and to measure the radio to X-ray offset of the pulses.

Cataclysmic variables (CVs) exhibit a wide range of timing
phenomena,including QPOs, X-ray transients, and complex light curves.
While CVs are typically $\sim 100$ times fainter than LMXBs, their
dynamical time scales are $\sim 1000$ times longer.  Moreover the
magnetic CVs, which will be USA's prime targets among CVs are
distinguished by having the largest magnetic moments among known
stellar populations, including even magnetars.  Curiously,
accretion-induced QPOs were predicted historically to result from this
flow before they were observed.  The QPOs have been seen repeatedly in
optical wavelengths but never in X-rays, despite searches.  Highly
correlated optical and X-ray luminosity variations are predicted in
current hydrodynamic models (Wolff, Wood \& Imamura 1991)


Finally, USA can exploit its great flexibility to observe targets of
opportunity that are deemed important by the science working group.
Already, USA has observed Aql X-1, the Rapid Burster, and X1630-472
during outbursts.  Of course, the accreting millisecond pulsar
SAXJ1808.4-3658 would be of particular interest if it becomes active
during the USA mission.

\begin{figure}
\centerline{\psfig{file=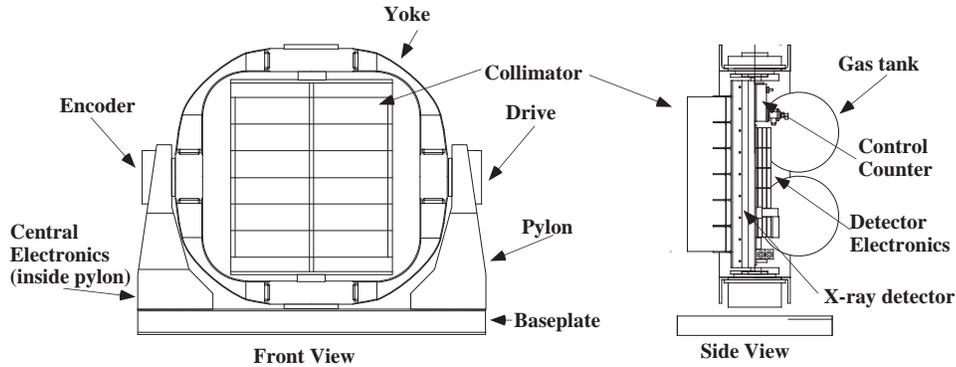, width=5in}}
\caption[]{Two Views of the USA Experiment}
\end{figure}

\section{Instrument Description}

\subsection{Proportional Counter X-ray Detectors}

The detector (Table~1) consists of two multiwire constant
flow proportional counters equipped with a 5.0~$\mu$m Mylar window
and an additional 1.9~$\mu$m thick aluminized Mylar heat shield. The
detector is filled with a mixture of 90\%\ argon and 10\%\ methane
(P-10) at a pressure of 16.1 psia (at room temperature). The detector
interior contains an array of wires which provides two layers of nine
2.8~cm square cells, each containing one anode wire, running the
length of the counter. An additional wire runs around the periphery of
the array as part of the cosmic ray veto system. The electronics are
designed to accept primarily X-ray events, which arise in one cell
only. Events registered in two or more wires by a cosmic ray track are
vetoed with an efficiency of about 99\%.

The high voltage on the anode wires is adjusted continuously to
stabilize the gain, using a feed-back loop which monitors the
pulse-height distribution of X-ray events in a small separate
proportional counter. Two discriminators provide a normalized value
independent of the absolute source intensity of the $^{55}$Fe source
used in the feed-back counter.

\begin{table}
\caption{Operational Features of the USA Detector System}
\begin{center}
\begin{tabular}{rl}
\hline
                   Gas: & P-10 at 1.1 atmosphere \\
             Flow rate: & 0.1 cc/min allows for 3--6 year  life \\
                Window: & 5.0 $\mu$m Mylar + $\sim$ 30--40\AA Nichrome\\
          Energy range: & 1--15 keV  \\
         Field of view: & collimation of 1.2\degree\ $\times$ 1.2\degree\ (FWHM) \\
     Energy resolution: &  0.17 (1 keV @ 5.9 keV), 128 raw PHA channels \\
  Aperture (effective): & 2000 cm$^2$ @ 3 keV \\
  On-orbit calibration: & solenoid operated $^{55}$Fe source \\
\hline
\end{tabular}
\end{center}
\end{table}

The collimators serve to support the window as well as to define the
field of view. To place reasonable requirements on the pointing
system the collimator was constructed with a field of view of
approximately 1.2\degree$\times$1.2\degree\ and a flat top of
approximately 0.05\degree. Each collimator consists of 8 modules 7.5
cm $\times$ 28 cm $\times$ 11 cm high filled with copper hexcell
formed from 25 $\mu$m sheet stock with a 2.5 mm altitude for each
hexagon. The sides and ends of each module are formed from 1 mm Cu
sheet which provides stiffness across the width of the collimator to
support a 98\% transmission nickel mesh that in turn supports the
Mylar window. The response function of each collimator module was
measured with X-rays before the modules were assembled into the
collimator frames.

\subsection{Support Hardware}

The ARGOS spacecraft is three-axis-stabilized and nadir-pointed. The
X-ray detectors are mounted on a 2-axis gimballed platform to permit
inertial pointing at celestial objects. The pointer is configured as
an equatorial mount looking aft (away from the velocity vector of the
spacecraft). Pointing is accomplished by a yaw rotation to acquire the
target followed by a continuous slew in the pitch to track the target
as it rotates about the orbital pole.

\begin{table}
\caption{Pointing System Characteristics}
\begin{center}
\begin{tabular}{rl}
\hline
     Pitch/Yaw drive capability: & $\sim 3.6$\degree /min (track),
	    $\sim 20$ \degree/min (slew) \\
		 Field of Regard: & 2 $\pi$ Sr \\
   Common pitch/yaw pivot design: & 180\degree\ travel in each axis \\
           	    Drive system: & 1.8 deg stepper motor \\
                            & 800:1 gear reduction, $\sim 20$ \arcsec/step \\
	 Position sensing system: & 16-bit optical shaft encoder \\
\hline
\end{tabular}
\end{center}
\end{table}

The primary tasks of the central command and control electronics (CE)
are command and data interface to the ARGOS spacecraft MIL-1553
bus, data acquisition from the detector modules, control of the pointer
system, and interface with USA's RH3000 and IDT3081 processors. The
command and control processor is a Harris radiation hardened 80C86
microprocessor.

The structural elements of the pointer (Table~2), the support pylons,
and the yoke which serves as the inner gimbal form the primary
structure of the experiment.  Each axis has a drive unit which forms
the pivot on one end of the axis and a position encoder unit which
supports the opposite end.  Actual alignment is measured by rastering
through sources in flight.

The detector interface board (DIB) in the CE performs time tagging and
data formatting for the X-ray science data as well as formatting
detector housekeeping data. The microprocessor used is an Analog
Devices ADSP2100. The DIB receives a fast photon arrival signal from
each detector which enables the timing to $\sim 1$ microsecond
accuracy.  A 1 Hz clock (with corresponding GPS time tag) is received
directly from the spacecraft to synchronize the event time tagging
clock.  Pulse height data for each photon are transmitted from the
detector to the DIB upon completion of the analog to digital
conversion. There are two standard telemetry modes: event and
spectral.  Event mode is the ``workhorse'' telemetry mode for USA;
for moderate count rates, it allows the maximal amount of information
to be preserved on each photon. In event mode, the arrival time and
some energy information is stored for each photon detected. There are
two submodes of event mode providing 32 $\mu$s time and 16 pulse
height channels in a 12 bit word and 2 $\mu$s time with 8 pulse height
channels in a 15 bit word respectively. Data may be output in event
mode at either 40 or 128 kbps providing maximum count rates of 3060 or
9940 events per second for 32 $\mu$s time or 2448 or 7952 events per
second for $\mu$s time. In spectral mode, a full resolution energy
spectrum (48 channels) is generated every 10 milliseconds for each
detector.

The USA experiment also provides space for two ``ride-along''
processor boards, the RH3000 and the IDT3081. The RH3000 board is
built around a pair radiation hardened Harris Semiconductor version of
the MIPS R3000 configured as a shadow pair with 2 MB of memory. The
IDT3081 board incorporates the commercial-off-the-shelf
IDT3081 processor and 2 MB of DRAM without any special error correcting
hardware.  Both computer boards have access to the downlink science
telemetry stream. These processors will be used to conduct experiments
in fault-tolerant computing, autonomous spacecraft navigation, and to
perform special data analysis functions which are beyond the scope of
the normal science telemetry modes, or which require bandwidths
greater than 128 kbps.

\subsection{Instrument Status}

The USA instrument has been performing well since activation began on
30 April 1999, but the USA mission has not been without its
difficulties. Approximately two weeks after launch the detector heat
shields suffered from degradation which has imposed additional
constraints on USA pointing with respect to the Sun. On 8 June 1999
Detector 2 suffered an event which increased the gas leak rate to a
very high level and exhausted the P-10 supply leaving only Detector 1
to complete the mission, halving the effective area. Two spacecraft
performance issues which are described in more detail below have also
impacted USA operations.

\section{ARGOS Mission Description}

USA exploits the flight opportunity provided by the ARGOS mission
under the DoD Space Test Program (STP). STP was established in
1965 as an activity under the executive management of the Air Force
Systems Command with the objective of providing spaceflight for DoD
research and development experiments which are not authorized to fund
their own flight. Both engineering/technology development and
scientific payloads have been flown with great frequency under this
program. ARGOS is the only Delta-class STP free-flyer mission to be
launched in the 1990s.

The 5000 lb ARGOS satellite was launched from Vandenberg AFB at 10:30
UT on 23 February 1999 aboard a Boeing Delta-II rocket.  The prime
satellite contractor was Boeing who built and tested the satellite at
their Seal Beach, CA facility.  ARGOS carries a complement of 9
experiments which address such topics as ionospheric remote sensing,
space dust, advanced electric propulsion, and high temperature
superconductivity.  

Spacecraft downlink telemetry bandwidth is done at 1, 4, or 5 Mbps.
Data are stored in a 2.4 Gbit solid state recorder and downlinked at
station passes to AFSCN ground stations.  The spacecraft is operated
in a 3-axis stabilized mode, with Z-axis of the spacecraft always
pointed to nadir.  Attitude control is based on a system of gyros and
horizon sensors feeding into reaction wheels and CO$_2$ thrusters. The
orbit is nearly circular with a 830 km altitude and a 98.7\degree\
inclination. It is Sun-synchronous with a beta-angle of 25--45
degrees, i.e., it crosses the equator at approximate local times of
14:00 on the day side and 02:00 on the night side.  This nearly polar
orbit means that USA encounters a high radiation environment multiple
times per orbit as it passes through the Earth's radiation belts at
latitudes above 50\degree.  This forces USA to take data at lower duty
cycle, turning off the detectors in the radiation belts and the South
Atlantic Anomaly to prevent detector breakdowns.

\subsection{Mission Operations and Data Processing}

The satellite mission operations are handled by the Air Force SMC/TEO
at Kirtland AFB, NM.  They are responsible for uplinking commands to
and for receiving data from the satellite.  Individual experiment
command uploads are delivered to TEO via FTP and uploaded during
ground contacts.  Data downlinked during the pass is recorded at the
ground station and mailed to TEO because the AFSCN does not support
real time links of $>1$ Mbps.  This results in a delay of 7--21 days
in getting science data back to the experimenters.

USA operation is largely automatic.  Twice daily command uploads
contain timed execution commands to slew the insrument, command it to
track the source, switch on the high voltage, select telemetry rate
and perform calibrations.  These command sequences are generated by a
highly automated observation scheduling system which optimizes source
selection, manages solid state recorder space, and builds the command
uploads.

The USA data processing system is also highly automated.  As data
appear at Kirtland, files are automatically retreived via FTP and the
first several processing steps are performed.  Quicklook data are
checked for anomalous conditions and the USA team is alerted by e-mail
if problems occur.  Subsequently, observations are extracted from the
Level 0 archive, converted to FITS, and distributed to the scientific
analysis centers, including NRL and SLAC.

A Science Working Group (SWG) has been established to help optimize
the scientific potential of USA. The SWG determines scientific
priorities for observing targets, subject to certain
constraints. Scheduling of targets during the USA mission will be
consistent with experiment science objectives, priorities, and mission
operations capabilities. Telemetry formats will be selected to support
overall objectives. USA has the flexibility to respond quickly to some
targets of opportunity with approximately a 1--3 day turnaround after
the decision to revise the observing plan. The SWG decides whether to
respond to potential targets of opportunity and also identifies
instances when coordinated observations with ground-based
observers, CGRO, RXTE, or other ARGOS instruments are scientifically
advantageous.  The USA team is receptive to collaborations to make
better use of the data, but the small size of the group does not allow
us to operate a conventional guest observer facility.

\subsection{ARGOS Mission Anomalies and Events}

The ARGOS launch and deployment went flawlessly, but since then
several problems have surfaced with various subsystems.  Generally,
the spacecraft has been very robust and has autonomously safed itself
when presented with dramatic disturbances, such as a battery exploding
on the electric propulsion experiment and during the USA Detector 2
gas leak.  Here we will just summarize the issues and describe how
they affect the operation of USA.

Shortly after launch, it was discovered that the GPS receiver is
unable to stay locked on to the GPS solution and provide good
navigation information.  This problem was traced to an unexpectedly
large input level to the receiver which causes cross correlation
errors which disrupt the solution.  Generally the receiver will lock
on, then oscillate between navigation and acquisition mode for a
period of a few minutes to a few hours before losing the solution
completely.  To recover the time resolution required for many of the
USA objectives, new software was uploaded to the satellite to make it
safer to initialize the receiver repeatedly.  Currently the receiver
is initialized 4 times per day.  Software is being developed to be
able to interpolate times using the onboard clock to recover precise
absolute times between periods when the receiver is locked on to a
good solution.

A problem was discovered with the Scanning Horizon Sensors which are
used to control the pitch and roll of the satellite.  They are more
radiation sensitive than expected and experience data dropouts or
return incorrect data during most passages through the South Atlantic
Anomaly (SAA).  This causes the spacecraft to respond and produces
attitude disturbences in the satellite when it is in the SAA.  This
does not affect USA because USA never operates in the SAA.


At the time of the software upload to work around the GPS receiver
problem, a problem with the offset pointing of USA from the satellite
appeared.  It appears that the navigation message sent to USA every
second no longer represents the true attitude of the spacecraft.  It
appears, after numerous scanning observations using USA, that the
satellite is out of alignment in the roll direction by about 1\degree.
The cause of this is currently unknown, and work is ongoing to
troubleshoot this problem and design a workaround.

\end{document}